\documentclass[12pt]{article}
\usepackage{arxiv}

\usepackage[utf8]{inputenc} 
\usepackage[T1]{fontenc}    
\usepackage{hyperref}       
\usepackage{url}            
\usepackage{booktabs, threeparttable}       
\usepackage{amsfonts, amssymb, amsmath, latexsym, amsthm, bm}       
\usepackage{nicefrac}       
\usepackage{microtype}      
\usepackage{lipsum}		
\usepackage{graphicx}
\usepackage{natbib}
\usepackage{doi}
\usepackage{pdflscape}
\usepackage{listings}
\title{Multidimensional constructs and moderated linear and nonlinear factor analysis}
\author{ \href{https://orcid.org/0000-0002-9114-3896}{\includegraphics[scale=0.06]{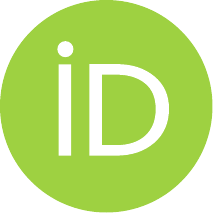}\hspace{1mm}R. Noah Padgett} \\
	Department of Epidemiology\\
	Harvard T.H. Chan School of Public Health\\
        Harvard University\\
	Boston, MA  \\
	\texttt{npadgett@hsph.harvard.edu}\\
}

\renewcommand{\shorttitle}{Multidimensional MNLFA}

\hypersetup{
pdftitle={Multidimensional constructs and moderated nonlinear factor analysis},
pdfsubject={stat.ME},
pdfauthor={R. Noah Padgett},
pdfkeywords={Bayesian, SEM, factor analysis, item response theory, differential item functioning, MNLFA},
}

\begin{document}
\large
\maketitle

\begin{abstract}
Multidimensional factor models with moderations on all model parameters have so far been limited to single-factor and two-factor models. This does not align well with existing psychological measures, which are commonly intended to assess 3-5 dimensions of a latent construct.  The methods for inducing moderated factor correlations are described including a method based on a partial correlations parameterization of Cholesky factors and a relative new method based on an iterative generalized Fisher transform. I conclude by discussing the theoretical implications of penalization for measurement invariance, computational considerations, and future directions for extending the framework to categorical indicators, longitudinal data, and applied research contexts.
\end{abstract}


\section*{Introduction}

A foundational component of nearly all scale development processes is investigating measurement invariance across subpopulations. 
Invariance of scale properties is sometimes referred to as fairness in an educational setting, as defined by the Standards (AERA, APA, NCME, 2024, p. 50) or to provide validity evidence for the consistency in interpretation of scores across groups of like respondents.
Many lines of research have developed to empirically assess the validity evidence supporting (or against) the conclusion that scores are comparable across groups \citep{Van_Erp2018, van-de-Vijver2019-ar}.

One way validity evidence can be provided is by evaluating the equivalence of sets of parameters in a confirmatory factor model across two or more groups (e.g., is the entire factor loading matrix equal for males and females? Or, are all item residual variances equal over time?).
Researchers have provided some excellent guidance and recommendations for scale developers to use, building on this perspective \citep{Vandenberg2000-vo}.
However, these methods can be clunky when trying to identify partial invariance, which is the likely scenario in which only a subset of parameters vary meaningfully across groups.

The degree of the non-invariance within subsets of parameters can be of substantive interest depending on the pattern of how parameters vary.
For example, in the study of psychopathology, indicators of internalizing tend to cluster into a few types of behaviors which are treated as factors to study different indicators of psychopathology for different types of people.
The pattern of how indicators of psychopathology vary across groups (e.g., sex and age) has a meaningful interpretation for the development of targeted interventions.
A limitation of the traditional factor analytic approach is the difficulty in identifying these patterns due to the invariance being studied from a set of parameters perspective.

An alternative perspective, often called differential item functioning, focuses on individual indicators (or items) instead of groups of parameters and has been a useful approach for identifying potentially variant indicators across groups \citep{Clauser1998-tp}.
Similarities with the multiple group factor analytic approach are discussed at length elsewhere \citep{Thissen2025-dw}.
A focus on indicators has been the predominant perspective in educational testing, aiming to identify and remove items that may be biased against specific student subgroups.
However, in psychological assessment, the degree of variance in parameters across groups can be of direct interest, as discussed in the study of psychopathology above.
A general approach that has gained attention in recent years for the study of invariance of model parameters across groups is known as moderated nonlinear factor analysis, which is a flexible framework for incorporating a design-like matrix of group/person characteristics to be predictors of factor analysis model parameters \citep{Bauer2023-hw}.

Moderated nonlinear factor analysis has been applied in many settings to identify variance in measurement model characteristics across several background characteristics simultaneously \citep{Bauer2023-hw}.
Other applications sometimes aim to provide scores on factors that are adjusted for possible partial invariance \citep{Curran2014-jm}.
The development of these methods has expanded into evaluating measurement invariance in up to two constructs \citep{Bauer2023-hw}.
Such moderated nonlinear factor analysis models allow for the simultaneous investigation of differences in latent factor parameters alongside measurement model characteristics.
Moderation effects on latent factors are sometimes referred to as the \textit{impact} of background characteristics on the construct of interest.
Investigating impact, while controlling for possible non-invariance of a small number of measurement model parameters, is of primary substantive interest when applying moderated nonlinear factor analysis.

The applications of moderated nonlinear factor analysis have been limited thus far to only two latent factors due to the inherent complexity of ensuring the latent variable covariance matrix is valid for all subgroups implied by the design matrix of moderating factors.
The current approach for two-factor models is to induce on an unbounded continuum and then use Fisher's z-transformation to restrict the parameter to the [-1,1] bound of correlations.
Extending to more than two latent factors cannot utilize the same method using independent unbounded scores transformed to the [-1,1] scale because independent applications of Fisher's z-transformations do not ensure the entire correlation matrix is valid.
Any methods for inducing impact on individual correlations must simultaneously ensure each correlation is between [-1,1] and that the matrix of correlations is positive definite.
Achieving these simultaneous constraints has been a longstanding problem in the simulation of correlation matrices  \citep{Joe2006-cl, Pourahmadi2015-hn, Archakov2021-ap} and in covariance prediction \citep{Pinheiro1996-tj, Pourahmadi1999-kj, Pourahmadi2011-se, Bucci2022-cv}.
Modified conceptualizations of these methods are used in this article to address the dimensional limitation of moderated nonlinear factor analysis. 
I generalize moderated nonlinear factor analysis to flexibly allow moderation in the latent factor correlation parameters in models with three or more factors.

The remainder of this article is organized as follows. 
Next, I describe the MNLFA model in more detail for one- and two-dimensional factor models.
Then, I described how MNLFA can be extended to $m$-dimensional factor models, which is the primary contribution of this article.
I provide examples of how the general specification can induce a variety of correlational structures for specific subgroups.
I then report on an application of the described multidimensional moderated nonlinear factor model for the study of psychopathology.
I end with a discussion of the model and potential limitations and areas of future study.

\section*{Moderated (non)Linear Factor Analysis}

I focus this article on latent variable models with continuous indicators, but these methods generalize to binary and Likert-type responses with some modifications.
To begin, consider the measurement model for items $i=1,\ldots,I$ with observed factor indicators $y$. 
The single factor latent variable $\eta$ for $n=1,\ldots,N$ cases can be described as
\begin{align}
y_{ni} &= \nu_{ni} + \lambda_{ni}\eta_{n} + \varepsilon_{ni}\\
\nu_{ni} &= \nu_{0i} + \mathbf{x}_n^t\bm\delta_{\nu i}\\
\lambda_{ni} &= \lambda_{0i} + \mathbf{x}_n^t\bm\delta_{\lambda i}\\
\varepsilon_{ni} \vert \mathbf{x}_n &\sim N\left(0.0, \theta_{ni}\right)\\
\theta_{ni} &= \theta_{0i}\exp(\mathbf{x}_n^t\bm\delta_{\theta i})\\
\eta_{n} \vert \mathbf{x}_n &\sim N\left(\alpha_n, \phi_{n}\right)\\
\alpha_{n} &= \alpha_0 + \mathbf{x}_n^t\bm\delta_{\alpha}\\
\phi_{n} &= \phi_0\exp{\left(\mathbf{x}_n^t\bm\delta_{\phi}\right)}.
\end{align}
A full list of the meaning of all parameters is provided in the appendix.
The major difference between the above model and a traditional single-factor model is the inclusion of moderation effects on model parameters (e.g, $\nu_{ni} = \nu_{0i} + \mathbf{x}_n^t\bm\delta_{\nu i}$ is the moderation on item intercepts).
One way to conceptualize this approach to parameterizing the factor model is to think of hierarchical linear models where the intercepts and slopes (factor loadings) are outcomes of interest \citep[see][p. 80]{Raudenbush2002-nn}. 
Shifting to a hierarchical/multi-level perspective for these models makes a direct connection between testing for non-invariance across a wide range of possible variables with well-known methods of linear modeling.

A major benefit of connecting evaluating non-invariance to linear modeling is the flexibility that allows for unique testing of potentially nonlinear relationships between person factors and non-invariance in parameters.
For example, \citet{Molenaar2021-mk} illustrated how generalized additive models could be used to evaluate non-invariance.
This flexibility opens the door to evaluating even more complex interrelationships between model parameters and person characteristics through interactions among characteristics.
Along with the benefits, evaluating non-invariance now shares many of the same concerns that arise with respect to collinearity in the design matrix for the predictors and overall model complexity.
I will return to discuss collinearity and model complexity when I discuss priors later.

The one-dimensional factor model described above is a valuable baseline for extending to more complex factor structures. 
The simplest extension is to a two-factor model with simple structure.
In such a two-factor model, the only new wrinkle lies in evaluating impact on the factor correlation.
\citet{Bauer2017-gs} suggested the use of the Fisher's z-transformation of a unconstrained parameter to handle the bounding constraint of correlations after the factor variances have been separated out.
\begin{align}
\begin{bmatrix} \eta_{ni} \\ \eta_{n2}\end{bmatrix} &\sim \text{MVN}\left(\begin{bmatrix} \alpha_{n1} \\ \alpha_{n2}\end{bmatrix}, \begin{bmatrix} \phi_{n11} & \phi_{n12}\\ \phi_{n12} & \phi_{n22}\end{bmatrix}\right)\\
\phi_{n11} &= \phi_{011}\exp{\left(\mathbf{x}_n^t\bm\delta_{\phi 1}\right)}\\
\phi_{n22} &= \phi_{022}\exp{\left(\mathbf{x}_n^t\bm\delta_{\phi 2}\right)}\\
\rho_{n12} &= 1 - \frac{2}{\exp{\left[2(\gamma_{012} + \mathbf{x}_n^t\bm\delta_{\gamma 21})\right]+1}}\\
\phi_{n12} &= \rho_{n12} \sqrt{\phi_{n11}}\sqrt{\phi_{n22}}
\end{align}
Fisher's z-transformation (Eq. 12) provides a straightforward method for inducing varying factor correlations separately from differences in factor variances.
However, to extend the model to three or more dimensions in the latent factor structure, Fisher's z-transformation approach does not work to jointly ensure the positive definiteness of the latent variable covariance matrix.
Next, I introduce an approach to ensuring the positive definiteness that could be used in m-dimensional latent variable space.

\section*{Multiple latent factors}

Extending beyond two latent factors effectively is a complex balancing act of \textit{where} in the model do you allow moderation to be estimated.
A fully saturated multidimensional model has moderation in all parameters--factor loadings, item intercepts, item residual variances, factor means, factor variances, and factor correlations.
However, as I discuss in more detail in the \textit{Estimation \& Empirical Identification} section, the fully saturated model may not be estimable in all situations due to limitations of the data and model at hand.
But first, I describe the fully saturated model and the process for inducing moderation in the factor correlation matrix to illustrate the potential for these models to explore a wide range of possible sources of moderation in different parts of the model.
Moderation in the item parameters is relatively straightforward and can be shown as below:
\begin{align}
y_{ni} &= \nu_{ni} + \bm\lambda_{ni}\bm\eta_{n} + \varepsilon_{ni}\\
\nu_{ni} &= \nu_{0i} + \mathbf{x}_n^t\bm{\delta}_{\nu i}\\
\lambda_{nim} &= \lambda_{0im} + \mathbf{x}_n^t\bm{\delta}_{\lambda im}\\
\varepsilon_{ni} \vert \mathbf{x}_n &\sim N\left(0.0, \exp(\theta_{0i} + \mathbf{x}_n^t\bm{\delta}_{\theta i})\right)\\
\bm\eta_{n} \vert \mathbf{x}_n &\sim MVN\left(\bm\alpha_n, \bm\Phi_{n}\right)\\
\bm\Phi_{n} &= \left(\bm\Phi^{D}_{n}\right)^{0.5}\left(\bm\Phi^{R}_{n}\right)\left(\bm\Phi^{D}_{n}\right)^{0.5}\\
\phi^{D}_{nm} &= \exp(\phi^d_{0m} + \mathbf{x}_n^t\bm{\delta}_{\phi d m})\\
\bm\Phi^{R}_{n} &=\ \ ?
\end{align}
The moderation in the item parameters is achieved through a model specification similar to that of the single or two-factor model.
A new wrinkle involves specifying how moderation is defined in the factor-loading matrix to account for potential cross-loadings.
Simple structure can be assumed where each item loads onto only one factor, or any number of cross-loadings can be specified.
The simplest solution is to fix all non-theorized cross-loadings to zero.
A more general solution is to freely estimate a subset of loadings which I believe primarily load on different factors, and then use regularizing priors on cross-loadings.
The complexity of extending to greater than two factors arises with the person-specific factor covariance matrices ($\bm\Phi_{n}$). 

Inducing moderation in the person-specific factor covariance matrices is the core limitation of multidimensional moderated nonlinear factor analysis.
This problem can be restated as the more general issue of predicting several covariances simultaneously \citep{Anderson1973-eb, Pourahmadi1999-kj}.
To simplify the problem, the factor covariance matrix can be decomposed into the factor variances (or standard deviations) and the factor correlations.
The problem now simplifies somewhat to how to construct a valid correlations matrix ($\bm\Phi^R_{n}$) from a set of possible moderated and unconstrained parameters.

Once the factor variances are separated out into a diagonal matrix of the variances ($\bm\Phi^D_{n}$).
Moderation in the factor variances can be easily induced using the exponential transform similar to the item residual variances.
The factor correlation matrix is the remaining piece, which I present an approach based on (1) a partial correlations and the Cholesky factorization, and (2) an iterative generalized Fisher transformation; each may be useful, though each also have limitations.

\subsection*{Partial correlations and the Cholesky factorization}
 
Cholesky factors serve as a useful intermediary step between an unconstrained space and a valid correlation matrix.
A valid Cholesky factor of a correlation matrix can be constructed using a partial correlation parameterization \citep{Joe2006-cl, Lewandowski2009}.
First, a vector of unconstrained parameters, let's call this vector $\bm\gamma$, has support of any real number. This vector, $\bm\gamma$, is then mapped to the constrained space of ``partial correlation'' [-1,1] using Fisher's z-transformation, then a Cholesky factor is built assuming the vector is comprised of partial correlations.
Moderating is induced in the first step on the unconstrained space, then the vector of moderated unconstrained parameters are mapped to partial correlations then the Cholesky factor is built.
This mapping from the unconstrained scale to the Cholesky factor is shown below:
\begin{align*}
    \bm\gamma &= [\gamma_{12},\gamma_{13},...,\gamma_{ij}]\\
    \gamma_{ij} &= \gamma_{0ij} + \mathbf{x}^t\bm\delta_{\gamma mm^\prime}\\
    \gamma^\ast_{ij} &= 1 - \frac{2}{\exp{\left[2(\gamma_{0ij} + \mathbf{x}^t\bm\delta_{\gamma ij})\right]+1}}\\
    \phi^L_{ij} &= \begin{cases}
        \hfil 0 &\text{if}\ j > i\\
        \hfil 1 &\text{if}\ 1 = j = i\\
        \hfil \gamma^\ast_{ij} &\text{if}, 1 = j < i\\
        \hfil \prod_{k=1}^{j-1} \left(1-\left(\gamma^\ast_{ik}\right)^2\right)^{0.5} &\text{if}, 1 < j = i\\
        \gamma^\ast_{ij} \prod_{k=1}^{j-1} \left(1-\left(\gamma^\ast_{ik}\right)^2\right)^{0.5} &\text{if}, 1 < j < i\\
    \end{cases}
\end{align*}
The Cholesky factor, $\bm\Phi^{L}$, can then be used to re-construct the factor correlation matrix
\begin{align}
    \bm\Phi^{R} &= \left(\bm\Phi^{L}\right) \left(\bm\Phi^{L}\right)^t.
\end{align}
The interpretation of each ``partial'' correlation ($\gamma^\ast_{ij}$) depends on the column and row of the Cholesky factor it maps to.
For instance, for $i=1,j=2$, $\gamma^\ast_{ij}$ is the correlation between factor 1 and 2, but element $i=3,j=2$ is the partial correlation between factors 2 and 3 after partialing out factor 1.
The resulting matrix $\bm\Phi^{L}$ can then be constructed for each person $n$ in a flexible way to induce different correlation structures depending on the design matrix ($\mathbf{x}_n^t$).
The above transformation is illustrated in a simple-to-use R function below.

\begin{lstlisting}
#' R Code
#' Builds a Cholesky factor (lower triangular matrix of dimension K) from a vector of unconstrained parameters (length K*(K-1)/2)
#' @param x, vector of unconstrainted partial correlations length(x) must equal K*(K-1)/2
#' @param K, dimension of matrix (KxK)
construct_chol_mat_lkj <- function(x,K){
  # validation check
  if(length(x) != K*(K-1)/2){
    warning(paste0("Number of supplied unconstrained partial correlations 'x' not valid for supplied 'K'. Length of x=", length(x),", but should equal K*(K-1)/2 = ",K*(K-1)/2))
  }
  z <- matrix(0, nrow=K,ncol=K) ;
  w <- matrix(0, nrow=K,ncol=K) ;
  r <- matrix(0, nrow=K,ncol=K) ;
  x_tanh <- numeric(length(x));
  x_tanh <- tanh(x); #hyperbolic tan tranform of y (unconstrained partial correlations)
  z[lower.tri(z,FALSE)] <- x_tanh;
  diag(z) = 1.0;
  for(i in 1:K){
    for(j in 1:K){
      if(i < j) w[i,j] = 0.0;
      if(1 == j && j == i) w[i,j] = 1.0;
      if(1 == j && j < i) w[i,j] = z[i,j];
      if(1 < j && j == i){
        w[i,j] = cumul_prod( (1.0 - (z[i,1:(j-1)])^2.0)^0.5 );
      }
      if(1 < j && j < i){
        w[i,j] = z[i,j] * cumul_prod( (1.0 - (z[i,1:(j-1)])^2.0)^0.5 );
      }
    }
  }
  w;
}
#' Compute cumulative product of a vector x
#' @param x a numeric vector (e.g., c(1,2,3,4))
#' @returns a numeric value of the resulting cumulative product of x.
cumul_prod <- function(x){
  K = length(x);
  z = 1;
  for(k in 1:K){
    z = z*(x[k]);
  }
  return(z);
}
#' Examples:
# Ex.1 2x2 matrix => a single correlation
corr_mat <- matrix(c(1,0.70,0.70,1.0),ncol=2)
corr_mat
x <- atanh(0.70) # inverse Fisher transform correlation to get input for cholesky construction
K <- 2
L <- construct_chol_mat_lkj(x, K)
L %*% t(L)
all.equal(L %*% t(L) , corr_mat) 
# Ex 2. 6x6 matrix
set.seed(1)
K <- 6; x <- rnorm(K*(K-1)/2) # generate a random set of partial cor.
L <- construct_chol_mat_lkj(x, K); L 
mycorr <- L %*% t(L)
det(mycorr) > 0 ## is the correlation matrix invertible?
\end{lstlisting}

\subsubsection*{Advantages: Computation}
A major benefit of the partial-correlations and Cholesky factorization approach is the well-known properties of the Cholesky matrix. 
It's easy to compute (as we show in the Reconstructing the covariance matrix section below).
One such property is that it can be used to parameterize the full covariance matrix for more efficient sampling of the multivariate normal distribution.
We show this in a non-centered parameterization of the Bayesian model implemented in Stan to estimate the MNLFA model.

\subsubsection*{Disadvantage: Factor order matters}
A potentially major confounding feature in the application of the proposed method for inducing moderated correlations is that the order in which factors are entered into the model determines which factors are partial out. This means that impact of a moderating variable on the correlation between factors 2 and 3 depends on any moderation in the heading correlations between factors 1 and 2 and factors 1 and 3.
Factor 1 being ``primary'' means the subsequent correlation moderation depends on the adequacy of factor one being identified.
More work will need to be done to unpack how severe the impact of factor ordering has on identification of moderation.
But, as an initial proposal, I would recommend ordering of the factors in increasing order of the degree of suspectibility to inter-factor correlation moderation. This means that we place the more "stable" factors first. Then, these more stable factors are partialed out when examining moderation in the residual correlation between subsequent factors.
This allows for more ``stable'' factors to be partialed out in the computation of later partial correlations, reducing the propagation of moderation effects through the ordering.

\subsubsection*{Reconstructing the Factor Covariance Matrix}
Recovering the covariance matrix from the unconstrained parameters is not exactly intuitive from the equations alone.
Suppose the inter-factor partial correlation are $-$0.50, 0.75, and 0.60. 
On the unconstrained scale (inverse hyperbolic tangent transform), these partial correlation are approximately $-$0.549, 0.973, and  0.693.
These partial correlations are mapped to the Cholesky factor of the implied correlation matrix by the following:
\begin{align*}
    P &= \begin{bmatrix}
        1 & 0 & 0\\
        -0.50 & 1 & 0\\
        0.75 & 0.60 & 1
    \end{bmatrix}\\
    &\text{Apply Transformation:}\\
    L &= \begin{bmatrix}
        1 & 0 & 0\\
        -0.50 & \left\lbrace 1 \times \left(1 - (-0.50)^2\right)^{0.5}\right\rbrace & 0\\
        0.75 & \left\lbrace0.60 \times \left(1 - (0.75)^2\right) ^{0.5}\right\rbrace & \left\lbrace 1\times \left(1 - (0.75)^2\right) ^{0.5} \times \left(1 - (0.60)^2\right) ^{0.5}\right\rbrace
    \end{bmatrix}\\
    L &= \begin{bmatrix}
        1.00 & 0 & 0\\
        -0.50 & 0.87 & 0\\
        0.75 & 0.39 & 0.53
    \end{bmatrix}
\end{align*}
therefore, the resulting factor correlation matrix is
\begin{align*}
    L L^t &= \begin{bmatrix}
        1.00 & -0.50 & 0.75\\
        -0.50 & 1.00 & -0.03\\
        0.75 & -0.03 & 1.00
    \end{bmatrix}
\end{align*}
With factor variances of 1.2, 1.5, and 0.80, the resulting factor covariance matrix is thus
\begin{align*}
   D L L^t D &= \begin{bmatrix}
        1.20 & 0.00 & 0.00\\
        0.00 & 1.50 & 0.00\\
        0.00 & 0.00 & 0.80
    \end{bmatrix}^{0.50}
    \begin{bmatrix}
        1.00 & -0.50 & 0.75\\
        -0.50 & 1.00 & -0.031\\
        0.75 & -0.031 & 1.00
    \end{bmatrix} 
    \begin{bmatrix}
        1.20 & 0.00 & 0.00\\
        0.00 & 1.50 & 0.00\\
        0.00 & 0.00 & 0.80
    \end{bmatrix}^{0.50}\\
    &= \begin{bmatrix}
        1.20 & -0.67 & 0.73\\
        -0.67 & 1.50 & -0.034\\
        0.73 & -0.034 & 0.80
    \end{bmatrix} 
\end{align*}

\subsection*{Iterative generalized Fisher transform}

Another approach for inducing moderated factor correlations is based on a relatively new method for building valid correlations matrices by \citet{Archakov2021-ap}.
Their approach to building a correlation matrix re-envisioned the problem from how to manipulate the off-diagonal elements into the form appropriate for a Cholesky factor to how to manipulate the diagonal elements of a matrix so that when the matrix exponential is a matrix with 1's along the diagonal.
A matrix exponential is defined as 
\begin{align*}
    e^{A} &= \sum_{k=0}^{\infty} \frac{A^{k}}{k!}
\end{align*}
where, $A^{k}$ is the element-wise power.
\citet{Archakov2021-ap} used the restrictive properties of a correlation matrix to show that the inverse of the matrix exponential, essentially the matrix logarithm can be used to iteratively update the matrix A until the exponential converges to a matrix with unit diagonals with all elements within the bounds of $(-1,1)$. (NOTE: \citet{Archakov2021-ap} mention that the general definition of a ``logarithm'' for a matrix is not a clean as their derivations show, but the restrictive case of a symmetric postive definite matrix makes the use of the ``logarithm'' feasible. Diving into these technical nuances is beyond the scope of the current work as it is sufficient for our purposes to have the foundation that the method is guaranteed to work for the case of correlations matrices.)
Of particular value from their work is the identification that the matrix exponential transform and inverse-transform is unique and no two sets of unconstrained values can lead to the same correlations matrix and vise-versa.
This is exactly what we need to induce moderated factor correlations. 
The authors called this the ``generalized Fisher transform'' hereafter shortened to ``GFT'' due to the similarities in the use of the concept of exponentiation under and unconstrained domain into a constrained space.

\begin{lstlisting}
#' R Code
#' Generalised Fisher Transform correlation matrix (Archakov & Hansen).
compute_cor_gft <- function(x, tol_value = 1e-4, max_iter = 1000) {
  if (!requireNamespace("expm", quietly = TRUE))
    stop("[compute_cor_gft] the expm package must be installed.")
  n <- 0.5 * (1 + sqrt(1 + 8 * length(x)))
  if (!(is.vector(x) && n %% 1 == 0)) stop("Dimension of 'x' is incorrect")
  if (!(tol_value >= 1e-14 && tol_value <= 1e-4)) stop("incorrect tolerance value")
  A <- matrix(0, nrow = n, ncol = n)
  A[upper.tri(A, diag = FALSE)] <- x
  A <- A + t(A)
  dvec <- diag(A); dist <- sqrt(n); niter <- -1L
  while (dist > sqrt(n) * tol_value) {
    diag_delta <- log(diag(expm::expm(A)))
    dvec <- dvec - diag_delta
    diag(A) <- dvec
    dist <- norm(matrix(diag_delta), type = "2")
    niter <- niter + 1L
    if (niter > max_iter) break
  }
  C <- expm::expm(A); 
  C
}

set.seed(314)
# Ex.1 3 factors
K<-3
myvalues <- rnorm(K*(K-1)/2)
mycor <- compute_cor_gft(myvalues)
det(mycor)
round(mycor,3)

# Ex.2 10 factors
K<-10
myvalues <- runif(K*(K-1)/2, -10,10)
mycor <- compute_cor_gft(myvalues)
det(mycor)
round(mycor,3)
\end{lstlisting}

\subsubsection*{Advantages: Independence of correlation}
The primary benefit of the iterative GFT approach is that moderation can be interpreted without the complex conditional language encountered by the partial correlation and Cholesky factorization approach.
Being able to interpret moderation in a specific correlation is a huge advance and removes the ordering issue of the partial correlations.

\subsubsection*{Disadvantage: It's iterative...}

Being an iterative approach to building the correlation matrix, makes this method potentially sensitive to the correlation structure, number of factors, and algorithm hyperparameters (such as number of iterations and convergence criteria).
While \citet{Archakov2021-ap} showed that the algorithm ``converged'' relatively rapidly (within 5-10 iterations) with all off-diagonal elements within $(-1,1)$ and all diagonal elements nearly 1.0, the \textit{how close are all the diagonals to 1.0?} is still a concern.
In exploring of the approach, I've encountered the algorithm also usually getting the diagonals very close 1.0 within 5-10 iterations, but occasionally takes 100-1000s of iterations when any of the off-diagonal elements of the unconstrained matrix are ``large.''
The way I've remedied this is to first pre-transform the unconstrained moderated factor correlations from $(-\infty, \infty) \rightarrow (-\pi,\pi)$ using a scaled hyperbolic tangent transform to pre-align the correlations closer to the origin making the iterative GFT converge more quickly.
This pre-conditioning step may not be necessary and is another layer of transformation, but it could help with computational speed, especially for models with many latent factors.

\begin{lstlisting}
K<-10
myvalues <- runif(K*(K-1)/2, -10,10)
myvalues.tanh <- pi*tanh(myvalues)
compute_cor_gft(myvalues )
compute_cor_gft(myvalues.tanh )
\end{lstlisting}

I've used $\pi$ as the upper limit on the pre-conditioned correlation to get the correlations effectively on a z-scale (approximately), but other values are certainly plausible and reasonable.
The arbitrary choice of $\pi$ effectively bounds the range of possible moderated correlations to $[-0.996, 0.996]$. 
Another equally arbitrary option is $3.80$ to effectively bound the correlations within $[-0.999, 0.999]$.
The bounding for the GFT is not exact given the iterative matrix exponential, but using the Fisher's Z-transformation is close.
It's not clear at this point how much this preconditioning helps or hurts and should be explored in greater detail to provide more exact guidance to help make the transform for efficient.

\subsubsection*{Updating in action}
Here, I show how the iterative GFT method works in practice using the R package expm to calculate the matrix exponential.
In the block below, I show the base unconstrained matrix which is not a correlation matrix (it has values > 1.0).
The updates to the matrix of unconstrained values and the implied correlation matrix are shown.

\begin{lstlisting}
# R Code + Output
> uvec <- c(1.2, 0.75, 1.33) # Unconstrained values
> fit_iters <- purrr::map(1:100, ~compute_cor_gft(uvec, max_iter = . - 2, tol_value = 1e-14))
> n <- 0.5 * (1 + sqrt(1 + 8 * length(uvec)))
> A <- matrix(0, nrow = n, ncol = n)
> A[upper.tri(A, diag = FALSE)] <- uvec
> A <- A + t(A)
> dvec <- diag(A); dist <- sqrt(n);
> # Iter 0: NOT A CORRELATION MATRIX
> print(A); 
# Unconstrained matrix
     [,1] [,2] [,3]
[1,] 0.00 1.20 0.75
[2,] 1.20 0.00 1.33
[3,] 0.75 1.33 0.00
# Implied Correlation matrix (clearly not even close)
> print(expm::expm(A))
         [,1]     [,2]     [,3]
[1,] 2.896808 2.965108 2.536197
[2,] 2.965108 3.744031 3.141967
[3,] 2.536197 3.141967 3.127786
> # Iter 1:
> diag_delta <- log(diag(expm::expm(A))) ; dvec <- dvec - diag_delta ## *** HERE IS THE CORE UPDATE *** - Shifts the diagonal by the previous iterations's log diagonal
> diag(A) <- dvec; A; print(expm::expm(A))
# Unconstrained matrix
         [,1]      [,2]      [,3]
[1,] -1.06361  1.200000  0.750000
[2,]  1.20000 -1.320163  1.330000
[3,]  0.75000  1.330000 -1.140325
          [,1]      [,2]      [,3]
[1,] 0.9512011 0.9050618 0.8056664
[2,] 0.9050618 1.0653705 0.9364487
[3,] 0.8056664 0.9364487 0.9754495
> # Iter 2:
> diag_delta <- log(diag(expm::expm(A)))
> dvec <- dvec - diag_delta
> diag(A) <- dvec
> A; print(expm::expm(A))
# Unconstrained matrix
         [,1]      [,2]      [,3]
[1,] -1.01358  1.200000  0.750000
[2,]  1.20000 -1.383485  1.330000
[3,]  0.75000  1.330000 -1.115469
# Correlation matrix
          [,1]      [,2]      [,3]
[1,] 0.9799910 0.9033885 0.8200378
[2,] 0.9033885 1.0310030 0.9270232
[3,] 0.8200378 0.9270232 0.9873593
> # Iter 10:
> expm::logm(fit_iters[[10]]); fit_iters[[10]]
# Unconstrained matrix
           [,1]      [,2]      [,3]
[1,] -0.9783304  1.200000  0.750000
[2,]  1.2000000 -1.441504  1.330000
[3,]  0.7500000  1.330000 -1.089638
# Correlation matrix - diagonals are still about +/- 0.0001 off
          [,1]      [,2]      [,3]
[1,] 0.9999673 0.8986670 0.8312536
[2,] 0.8986670 1.0000782 0.9189267
[3,] 0.8312536 0.9189267 0.9999545
> # Iter 100:
> expm::logm(fit_iters[[100]]); fit_iters[[100]]
# Unconstrained matrix
           [,1]      [,2]      [,3]
[1,] -0.9782687  1.200000  0.750000
[2,]  1.2000000 -1.441653  1.330000
[3,]  0.7500000  1.330000 -1.089551
# Correlation matrix after 100 iterations
          [,1]      [,2]      [,3]
[1,] 1.0000000 0.8986469 0.8312802
[2,] 0.8986469 1.0000000 0.9189116
[3,] 0.8312802 0.9189116 1.0000000
\end{lstlisting}

\subsection*{Impact of a continuous variable on correlations}

\begin{figure}[!htp]
    \centering
    \includegraphics[width=0.75\linewidth]{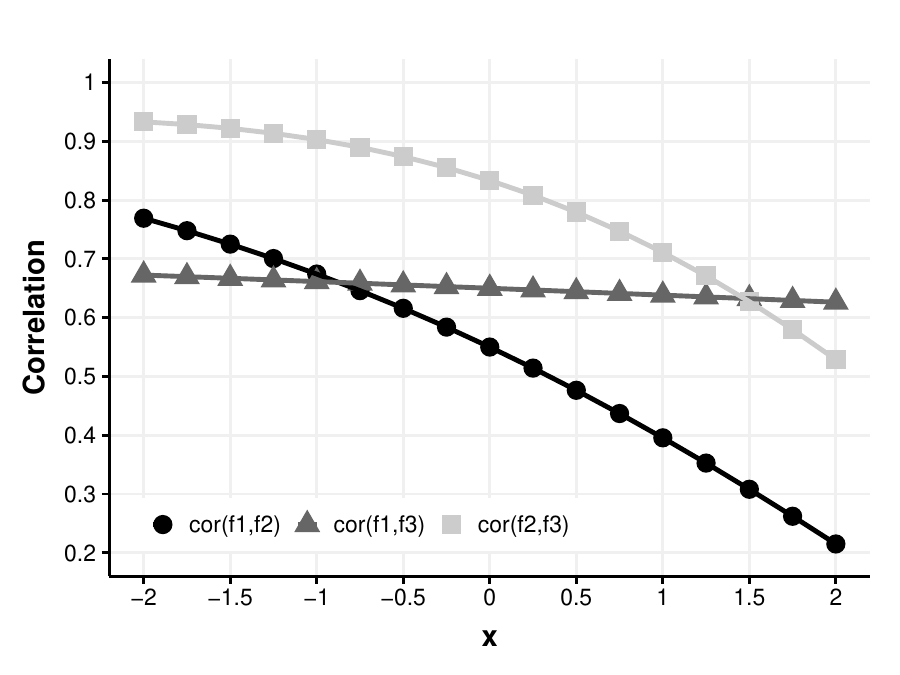}
    \caption{Induced varying correlations among three factors illustration differentiation among factors as a function of a continuous design factor $x$. Parameters used to simulate data based on the partial correlation parameterization: baseline residual correlations--$\gamma_{12}=0.55$, $\gamma_{13}=0.65$, $\gamma_{23}=0.75$;  magnitude of effects on $x$ on residual correlations--$\delta_{\gamma 12}=-0.20$; $\delta_{\gamma 13}=-0.02$; $\delta_{\gamma 23}=-0.20$.}
    \label{fig:vary-cor}
\end{figure}

The potential of this approach to inducing varying correlation structures depending on person parameters is shown in Figure \ref{fig:vary-cor}.
The figure plots the induced varying correlations across a continuous design variable $x$, where on average, the correlations are $\text{cor}(f_1, f_2)=0.55$, $\text{cor}(f_1, f_3)=0.65$, and $\text{cor}(f_2, f_3)=0.83$; but, as shown, the magnitude of correlation can vary quite substantially as a function of a continuous design factor.
Notably, although the effect of the predictor is linear in the space of the unconstrained parameters being moderated, the effect on the induced correlation is nonlinear due to the transformation.
The nonlinearity is more noticeable for the correlation between factors 2 and 3 ($\text{cor}(f_2, f_3)$) because impact is propagated through the impact already present for the other two correlations; that is, the impact on the correlation between factors 2 and 3 is a function of the impact on the residual correlation $\text{cor}(f_2, f_3)$ and the impact on the correlation between factors 1 and 3 ($\text{cor}(f_1, f_3)$). 

\subsection*{Determining significant of impact}
A complicated aspect of exploring the impact of moderating variables on latent correlations is how to determine if the impact is ``large enough'' to be meaningful for discussion or not.
This is especially important in the partial correlation-Cholesky factorization approach where the direct "impact" in terms of the size of the coefficient is not directly obvious, especially for correlations later down the chain.

\section*{Estimation \& Identification}

Estimating moderated nonlinear factor analyses is an active area of methodological research \citep{Brandt2025-aa, Chen2024-ko, Gottfredson-2019, Bauer2020-dw, Kolbe2024-tm}.
A common theme across these studies involved the use of regularization to help identify the model given the complexity of the likelihood.
An approach with promise involves using Bayesian methods to approximate the posterior distribution and regularizing priors such as spike-and-slab priors for the DIF parameters \citep{Brandt2025-aa, Chen2024-ko}.
The R package \citep{Belzak2023-kv} implements Bayesian moderated nonlinear factor analyses for models with up to two latent factors relying on regularizing priors for DIF parameters \citep{Belzak2023-kv}
regDIF uses the general Bayesian MCMC package Just Another Gibbs Sampler (JAGS) to sample the marginal posterior distributions.
JAGS is incredibly powerful and flexible, but can be painfully slow for models with large numbers of parameters.
To help overcome this limitation, I implemented the multidimensional moderated nonlinear factor analysis in Stan \citep{Carpenter2017}, which is theoretically more efficient and can more easily explore the complex posterior distributions.
I implemented the multidimensional MNLFA model in Stan (described next) and discuss issues encountered and remedies.

\subsection*{Bayesian approach}

A Bayesian latent variable model can be implemented in several different ways depending on how the likelihood is set-up.
Initially, I built the multidimensional MNLFA using a \textit{conditional} likelihood approach to bring the implementation in line with how the package 'brms' is likely to implement latent variable models in version 3 (\url{https://github.com/paul-buerkner/brms/issues/304}).

The likelihood is implemented as shown in Figure \ref{fig:msd-mulit-mnlfa}.
A complexity when implementing the model in Stan is the need to create independent foundation parameters that are then transformed into the parameters necessary for the likelihood function contribution itself.
Also known as a \textit{non-centered} parameterization so that the parameters being directly sampled have no direct dependence. 
This is accomplished largely by using independent normal parameters that are then transformed to get the quantities of interest.
For instance, assuming there are $N$ respondents and $M$ latent variables, then the non-centered parameterization for the latent factors can be implemented to theoretically increase efficiency.

The non-centered parameterization is very memory intensive for moderated item factor analysis (MNLFA with binary/ordinal items) due to the need to supply `nuisance'' parameters ($u$) to utilize the underlying latent response distribution of the probit model.
Alternatively, one could use a logistic regression-based approach to bypass this requirement, as done by \citet{Chen2024-ko}. 
Though, it is unclear how well this approach will extend to multidimensional constructs with complex covariance structures.
the non-centered parameterization for the latent factors can be implemented as:
\begin{lstlisting}
# Stan code
data {
    int N; //number of cases
    int P; //number of observed variables
    int M; //number of latent factors
    array[N] vector[P] y_obs;
}
parameters {
 array[N*M] real eta_raw;  //independent latent factors (MVN)
 ... //other parameters
}
transformed parametrs {
    array[N] vector[M] eta; //rescaled factor scores
    for(n in 1:N){
        //Alpha = moderated factor means (Mx1)
        //Psi_chol, moderated factor correlation matrix-cholesky parameterization (lower MxM)
        //Psi_var, moderated factor variances (Mx1)
        //Lambda, moderated factor loading matrix (PxM)
        //Nu, moderated item intercepts (Px1)
        //Theta, diagonal matrix of moderated item residual variances (PxP)
        .
        .
        .//code to get moderated parameters (omitted for simplicity)
        eta[n] = Alpha + to_vector((Psi_chol * to_vector(eta_raw[((n-1)*M+1):(n*M)]) )' * diag_matrix(Psi_var)); // non-centered parameterization of latent factor scores
        implied_mean[n] = Nu + Lambda*eta;
        implied_residual_cov[n] = Theta;
    }
}
model {
    for(n in 1:N){
        //to help conserve memory, the transformed parameters could be constructed within this loop instead of the transformed parameters block
        target += multi_normal_chol_lpdf(y[n] | implied_mean[n], implied_residual_cov[n]);
    }
    target += std_normal_lpdf(eta_raw);
    .
    .
    .//other priors
}
\end{lstlisting}

Implementing the \textit{conditional} likelihood parameterization had several issues.
First, it was PAINFULLY slow, partially due to the huge number of parameters (number of cases $\times$ number of latent factors).
This approach becomes even more memory intensive for item factor analysis (MNLFA with binary/ordinal items) due to the need to supply `nuisance'' parameters $u$ to utilize the underlying latent response distribution of the probit model.
To help overcome this limitation, I rewrote the model to utilize the \textit{marginal} likelihood, similar to what is utilized in lavaan \citep{lavaan} for estimating simple factor models, as I describe in more detail in the next section.
Secondly, the conditional likelihood approach led to computer memory issues.
When trying to implement the model above, I could not get the model to run with sample sizes above 5000 (individual cases, not individual responses even for "small" models with only 10-20 items) without requesting over 1000 GB of RAM on a cluster computer.
Due to these issues, I decided to try an alternative parameterization.

\subsubsection*{Marginal likelihood approach}

The marginal likelihood implemented in Stan is as follows.
If the observed factor indicators are continuous variables, the person-wise marginal likelihood can be expressed as
\begin{align}
\bm{y}_{n} &\sim \text{MVN}\left(\bm\nu_{n} + \bm\Lambda_{n}\bm\alpha_{n}, \bm\Lambda_n\bm\Phi_n\bm\Lambda_n^t + \bm\Theta_n\right)\\
\nu_{n,i} &= \nu_{0i} + \mathbf{x}_n^t\bm{\delta}_{\nu i}\\
\lambda_{n,im} &= \lambda_{0im} + \mathbf{x}_n^t\bm{\delta}_{\lambda im}\\
\theta_{0,i} &= \theta_{0i}\exp(\mathbf{x}_n^t\bm{\delta}_{\theta i})\\
\alpha_{n,m} &= \alpha_{0m} + \mathbf{x}_n^t\bm{\delta}_{\alpha m}\\
\bm\Phi_{n} &= \left(\bm\Phi^{D}_{n}\right)^{0.5}\left(\bm\Phi^{L}_{n}\right)\left(\bm\Phi^{L}_{n}\right)^t\left(\bm\Phi^{D}_{n}\right)^{0.5}\\
\bm\Phi_{n}^D &= \text{diag}(\phi_{n1},...,\phi_{nm})\\
\phi_{n,m} &= \phi_{0m}\exp(\mathbf{x}_n^t\bm{\delta}_{\phi m})\\
\bm\Phi_{n}^L &= f(\bm\gamma_n)\\
\gamma_{n,m^\prime} &= \gamma_{nm^\prime} + \mathbf{x}_n^t\bm{\delta}_{\gamma m^\prime}
\end{align}
The function $f(.)$ in equation 28 is the transformation from unconstrained partial correlations to the lower-triangular Cholesky factor of the factor correlation matrix.
Using this marginal likelihood framing ensures the dimension of the parameter space does not increase with sample size, leading to more efficient sampling.
A major benefit of this approach is that common SEM/CFA identification rules \citep{Bollen1989} can be used to ensure the baseline parameters are easily identified, thereby facilitating the empirical identification of the remaining parameters.
Examples of identification approaches for the baseline model are:

\textit{Fixed baseline factor scale}. A fixed baseline factor variance is also possible. Such as fixing the factor variance to 1.0, and allowing the factor variance to vary as a function of background characteristics. This can work well in conjunction with a semi-informative prior on the factor loadings to establish the scale of the factor variables while still allowing for population differences in scale as a function of background characteristics.
These differences would then be interpretable as changes in scales compared to the average across populations.

\textit{Marker variable/fixed loading}. Constraining the baseline factor loadings of each variable/item's primary factor loading to be a fixed value or sign-constrained.
This can be accomplished using a constraint on the parameter itself, or using a semi-informative prior that does not put much, if any, weight below zero; e.g., $\lambda_{0,im} \sim \text{Normal}(1.0, 0.5)$. 

\subsubsection*{Spike-and-slab priors for DIF \& Impact}

Spike-and-slab priors have seen a growing interest in the area of DIF detection \citep{Brandt2018-dg, Belzak2020-rj, Chen2021-ni, Brandt2025-aa}.
The spike-and-slab prior can be implemented in a variety of ways depending on the needs of the situation.
One way previously implemented is 
\begin{align}
\delta &\sim \omega \text{Normal}\left(0, \psi\tau\sigma^2\right)\\
\psi &\sim \text{exponential}\left(0.5\right)\\
\omega &\sim \text{Beta}\left(1,1\right)\\
\tau^{-1} &\sim \text{Gamma}\left(9,4\right)
\end{align}
where $\delta$ is a placeholder for any of the DIF/impact effects, $\omega$ is the inclusion probability, $\psi$ and $\tau$ are regularizing variance parameters across a group of jointly regularized parameters, and $\sigma^2$ is the item specific residual variance.

When implementing the above parameterization of the spike-and-slab prior, I found that the individual regularizing variance parameters were not unique leading to sampling difficulties in Stan. 
An alternative parameterization designed specifically for Stan is described next. 
\citet{Thomson2019-va} provided a logit-normal continuous parameterization that is especially efficient for use in Stan since all sampled parameters are continuous with unbounded support that can then be transformed to the standard spike-and-slab structure.
\begin{align}
    \delta &\sim \omega \times \text{Normal}\left(0, \tau\right)\\
    \text{logit}(\omega) &= \log\left(\frac{\omega_{\mu}}{1-\omega_{\mu}}\right) + \text{Normal}\left(0, 1\right)\times \omega_{sd} 
\end{align}
where $\omega$ is the inclusion probability with hyperparameters $\omega_{\mu}$ for prior average inclusion probability and $\omega_{sd}$ for the standard deviation of the logit of the inclusion probability, and $\tau^2$ is the prior standard deviation of the DIF/impact effect.
The above prior parameterization has $\tau$ fixed (e.g., $\tau = 5.0$), as done in all the simulations conducted by Thompson and colleagues, which greatly improves the speed and sampling efficiency of the spike and slab prior.
The standard deviation prior can itself be given a hyper-prior, such as a Gamma distribution, to provide an ``adaptive'' component \citep{Brandt2025-aa}.
However, in experimentation, I've found that including the ``adaptive'' piece leads to a significant increase in computation time and slightly more diffuse posterior distributions for the DIF/impact effects.
A major advantage of the fixed variance approach is only needing to use standard normal priors then shifting/rescaling as needed combined with the inclusion probability to achieve a more or less regularizing prior distribution; that is, changing the prior inclusion probability (mean, $\omega_\mu$, and standard deviation, $\omega_{sd}$) and increasing or decreasing $\tau$ for more or less regularizing.
Together, this leads to a relatively straightforward interpretation of the regularizing prior that is easy to specify and modify as needed for sensitivity analyses.

\begin{lstlisting}
# Stan code
data {
    int Q;//number of regularied parameters
    real pen_mu ; // pre-transformed IP mean on logit scale
    real<lower=0> pen_sd ;
    real<lower=0> pen_tau;//if a fixed value
}
parameters {
    array[Q] real delta_raw;
    real pen_phi;
    //array[Q] real<lower=0> pen_tau;
}
transformed parameters {
    real delta;
    real pen_inclusion_prob;
    pen_inclusion_prob = inv_logit(pen_mu + pen_sd .* to_vector(pen_phi));
    delta = detla_raw .* pen_tau .* pen_inclusion_prob;
}
model {
 ...
 target += std_normal_lpdf(delta_raw); //dif/impact raw parameter
 target += std_normal_lpdf(pen_phi); //inclusion probability
 //target += gamma_lpdf(pen_tau | tau_a, tau_b); //only used if allowing for varying penalty variance
}
\end{lstlisting}

\subsection*{Penalized maximum likelihood}

The severe computational time led described in the Bayesian section above, I decided to explore alternative estimation approaches and identifying where the bottleneck in computation is in estimating the model.
The reason the model in Stan was taking so long is that, under the hood, Stan is approximating the gradient, the first derivative of the posterior, for each model parameter, and then moving along this approximated curve to get to the next position in the posterior surface. 
This gradient approximation was very computationally demanding in a general multidimensional MNLFA model, leading to the excessive slowness I observed. 
To overcome this complexity, I sought an estimation approach that didn't require a numerical approximation of the gradient.
I returned to old school maximum likelihood methods.
Because the MNLFA model assumes a multivariate normal distribution for the responses, the likelihood function is the well-known multivariate normal, and as it turns out, the gradients for all model parameters in the MNLFA have a closed form that I can make use of when estimating the model.

Traditional methods of maximum likelihood do not incorporate regularization, but I can incorporate penalization for complexity that acts as a regularizing factor \citep{Asparouhov2024-ia}.
Instead of directly working with the log-likelihood, I can optimize a weighted composite of the log-likelihood ($L(.)$) and a penalty ($P(.)$),
\begin{align}
L_w(\theta) &= L(\theta) - w_0P(\theta) ,
\end{align}
where $w_0$ is the weight given to the "prior" or the penalty term, $\theta$ is the vector of all parameters being estimated, and $L_w(\theta)$ represents the composite log-likelihood.
The penalty term can take different forms, but the method I propose is based on a splitting the parameter vector into pieces $(\theta_0, \theta_1)$ where $\theta_0$ are the non-regularized terms (e.g., baseline item intercepts, factor loadings, etc.) and where $\theta_1$ contains the DIF/impact parameters. 
The penalty is then placed only with respect to $\theta_1$. 
A few such penalties can be defined by summing up all the squared differences among the parameters in $\theta_1$,
\begin{align}
    P(\theta_1) &= \sum_{\forall i} \sum_{\forall j \neq i} \frac{(\theta_i - \theta_j)^2}{\nu}, \text{the ridge penality}\\
    P(\theta_1) &= \sum_{\forall i} \sum_{\forall j \neq i} \frac{\sqrt{(\theta_i - \theta_j)^2 + \varepsilon}}{\nu}, \text{the LASSO penality}\\
    P(\theta_1) &= \sum_{\forall i} \sum_{\forall j \neq i} \frac{ \sqrt[4]{(\theta_i - \theta_j)^2 + \varepsilon} }{\nu}, \text{the Alignment penality}
\end{align}
where $\nu$ controls how severe the penalty for large differences among parameters in $\theta_1$. 
\citet{Asparouhov2024-ia} describe using the alignment penalty parameterized as $\sqrt{|\theta_i -\theta_j|+\varepsilon}$, which is approximate the same as shown above; however, using the quartic root is a smooth function and easier to differentiate and thus preferred for our purposes.

The penalized log-likelihood itself is relatively straightforward to set up to maximize using numerical methods, such as using routine optimization routines in any statistical package (e.g., optim(.) or nlminb(.) in \textbf{R}).
However, MNLFA requires using a person-wise constructed mean and covariance matrix when evaluating the multivariate normal density function. 
This creates a computationally complex situation where I cannot use sufficient statistics to simplify the overall likelihood expression, as is done in lavaan.
Instead must evaluate the density function for each case, or person, in our dataset after constructing the person-specific model implied mean and covariance matrix for the observed data.
This complexity issue is similar to those found when trying to set up routines to estimate a multilevel structural equation model using full information maximum likelihood \citep[see ][ for a deeper discussion on this topic]{Rosseel2021-sj}.
However, the tricks for efficiency described by \citet{Rosseel2021-sj} may not be able to improve the efficiency of evaluating the likelihood in MNLFA because no two persons in a dataset may have the exact same combination of moderating variables ($X$) or missing data pattern; though I certainly could be mistaken about this limitation and future work should seek explore such possibilities.
Due to this constraint, the only major source for efficiency gains in estimating the model lies with how the gradient of the log-likelihood is evaluated.

\subsubsection*{Gradient approximation}

As noted previously, evaluating the gradient of the likelihood was the limiting factor of Stan for estimating these models.
A numerical approximation of the gradient of the log-likelihood is also quite time-consuming, and the approximation increases in time needed with model size and sample size.
Due to this limitation of existing methods, I derived the expressions for the gradients to alleviate the need for complex approximations of the full gradient.

Using the marginal likelihood with respect to the latent factors, the expression for the person-wise likelihood contribution is

\begin{equation}
L_n = -\frac{1}{2}\ln{2\pi} -\frac{1}{2} \log|\bm\Sigma_n| - \frac{1}{2} (\mathbf{y}_n - \bm\mu_n)^\top \bm\Sigma_n^{-1} (\mathbf{y}_n - \bm\mu_n) .   
\end{equation}

Where $\bm\Sigma_n = \bm\Lambda_n \bm\Phi_n \bm\Lambda_n^t + \bm\Theta_n$ and $\bm\mu_n = \bm\nu_n + \bm\Lambda_n\bm\alpha_n$. The goal now is to derive the expression for the gradient (or partial derivatives) of this likelihood contribution with respect to each possible model parameter.
This step can further be broken down into the baseline values for each parameter type (e.g., baseline item intercepts and baseline factor correlations) and the moderation effects on these parameters (e.g., each $\bm\Delta$).
This wasn't as terrible to derive as I expected, given the well-known properties of the multivariate normal distribution. 

For simplicity, I drop the subscript with respect to person $n$, but I assume that all parameters are conditional at the person-level.
The gradient with respect to the baseline item intercepts $\nu$ are:
\begin{equation}
\frac{\partial L}{\partial \bm\nu} = \bm\Sigma^{-1} (\bm{y} - \bm\mu).
\end{equation}

\begin{lstlisting}
#' R Code
#' @param ri residual of observed scores minus implied mean (y - mu)
#' @param invSigma pre-computed invserse of implied covariance matrix
grad_nu <- function(ri, invSigma){
  invSigma %*% ri
}
\end{lstlisting}

The gradient for a specific item intercept is obtained by extracting the correct row from the resulting column vector of partial derivatives.
The gradient of the loglikelihood function w.r.t. to the factor loadings are a bit more involved due to the factor loadings contributing to the location and covariance matrices.
\begin{align}
\frac{\partial L}{\partial \bm\Lambda_0} &= \left( \frac{\partial \bm\mu}{\partial \bm\Lambda} \right)^t \Sigma^{-1} (\bm{y} - \bm\mu) + \frac{1}{2} \operatorname{tr}\left[ \left( \bm\Sigma^{-1} (\bm{y} - \bm\mu)(\bm{y} - \bm\mu)^t \bm\Sigma^{-1} - \bm\Sigma^{-1} \right)
\frac{\partial \bm\Sigma}{\partial \bm\Lambda_0} \right]\\
\frac{\partial \bm\mu}{\partial \bm\Lambda_0} &= \bm\alpha\\
\frac{\partial \bm\Sigma}{\partial \bm\Lambda_0} &= \bm\Phi \bm\Lambda^t + \bm\Lambda \bm\Phi
\end{align}
\begin{lstlisting}
#' R Code
#' @param i focal item
#' @param f focal factor of item i
#' @param ri residual of observed scores minus implied mean (y - mu)
#' @param invSigma pre-computed invserse of implied covariance matrix
#' @param Lambda implied factor loading matrix
#' @param Alpha implied factor means
#' @param Phi implied factor covariance matrix
#' @param Qsr pre-compute quadratic form of invSigma and residual (Qsr= (invSigma) (ri) (ri^t) (InvSigma) - invSigma)
grad_lambda <- function(i, f, ri, invSigma, Lambda, Alpha, Phi, Qsr){
  # 1. Dimensions
  p = nrow(Lambda)
  # 2. Derivative of the mean
  dmu <- matrix(0,ncol=p, nrow=1)
  dmu[1,i] = Alpha[f,1]
  grad_mean_part = as.numeric(sum(dia(dmu %*% (invSigma %*% ri))))
  # 3. Derivative of the covariance
  e_i <- matrix(0, nrow=p, ncol=1)
  V_f = Lambda %*% Phi[,f];
  e_i[i,1] = 1;
  dSigma = e_i %*% t(V_f) + V_f %*% t(e_i);
  grad_cov_part = 0.5 * as.numeric(trace(Qsr %*% dSigma));
  # Total gradient
  grad_mean_part + grad_cov_part
}
\end{lstlisting}
where again, the gradient is the specific element of the results matrix.
The gradient w.r.t. the item residual variances is
\begin{align}
\frac{\partial L}{\partial \bm\Theta_{0}} &= \frac{1}{2} \bm\Theta_{0} \left(\bm\Sigma^{-1} (\bm{y} - \bm\mu)(\bm{y} - \bm\mu)^t \bm\Sigma^{-1} - \bm\Sigma^{-1}\right)
\end{align}

\begin{lstlisting}
#' R Code
#' @param i focal item
#' @param Qsr pre-compute quadratic form of invSigma and residual (Qsr= (invSigma) (ri) (ri^t) (InvSigma) - invSigma)
#' @param Theta matrix of implied residual covariances
grad_theta <- function(i, Qsr, Theta){
  gx = 0.5 * Theta[i,i] * as.numeric( Qsr[i,i] )
  gx
}
\end{lstlisting}

The gradient w.r.t. the factor means is
\begin{align}
\frac{\partial L}{\partial \bm\alpha_0} &= \bm\Lambda^t \bm\Sigma^{-1} (\bm{y} - \bm\mu)\\
\frac{\partial \bm\mu_i}{\partial \bm\alpha_0} &= \bm\Lambda
\end{align}
\begin{lstlisting}
#' R Code
#' @param ri residual of observed scores minus implied mean (y - mu)
#' @param invSigma pre-computed invserse of implied covariance matrix
#' @param Lambda implied factor loading matrix
grad_alpha <- function(ri, invSigma, Lambda){
  gx = t(Lambda) %*% invSigma %*% ri
  gx
}
\end{lstlisting}

The gradient w.r.t. to the factor variances is
\begin{align}
\frac{\partial L}{\partial \bm\Phi^D} &= \frac{1}{2} \operatorname{tr}\left[ \left(\bm\Sigma^{-1} (\bm{y} - \bm\mu)(\bm{y} - \bm\mu)^t \bm\Sigma^{-1} - \bm\Sigma^{-1}\right) \cdot \frac{\partial \bm\Sigma}{\partial \bm\Phi^D} \right]\\
\frac{\partial \bm\Sigma}{\partial \bm\Phi^D} &= \bm\Lambda \left(\bm\Phi^L\right)^t \left( \frac{\partial \bm\Phi}{\partial \bm\Phi^D} \right) \left(\bm\Phi^L\right) \bm\Lambda^t
\end{align}

The gradient w.r.t. to the factor correlations is more involved due to the nested dependence of the gradients.
\begin{align}
\rho_{jk} &= \tanh(z_{jk} + \mathbf{x}^t\bm\delta_{jk})\\
\frac{\partial L}{\partial \bm\rho} &= \frac{1}{2} \operatorname{tr}\left[ \left(\bm\Sigma^{-1} (\bm{y} - \bm\mu)(\bm{y} - \bm\mu)^t \bm\Sigma^{-1} - \bm\Sigma^{-1}\right) \cdot \frac{\partial \bm\Sigma}{\partial \bm\rho} \right] \cdot \frac{\partial\bm\rho}{\partial\bm{z}}\\
\frac{\partial \bm\Sigma}{\partial \bm\rho} &= \bm\Lambda \left( \frac{\partial \bm\Phi}{\partial \bm\rho} \right) \bm\Lambda^t\\
\frac{\partial\bm\rho}{\partial\bm{z}} &= 1 - \tanh^2(\bm{z})\\
\frac{\partial \bm\Phi}{\partial \bm\rho} &= \frac{\partial L}{\partial \bm\rho} \bm\Lambda^t + \bm\Lambda \left( \frac{\partial L}{\partial \bm\rho} \right)^t\\
\frac{\partial L}{\partial \bm\rho} &= \begin{bmatrix}
    1 & 0 & \cdots & 0\\
    \frac{\partial L_{21}}{\partial \rho_{jk}} & \frac{\partial L_{22}}{\partial \rho_{jk}} & \cdots & 0\\
    \frac{\partial L_{31}}{\partial \rho_{jk}} & \frac{\partial L_{33}}{\partial \rho_{jk}} & \cdots & \vdots\\
    \vdots & \vdots & \ddots & \vdots\\
    \frac{\partial L_{r1}}{\partial \rho_{jk}} & \frac{\partial L_{r2}}{\partial \rho_{jk}} & \cdots &\frac{\partial L_{rs}}{\partial \rho_{jk}}
\end{bmatrix}, \text{where}\quad
\frac{\partial L_{rs}}{\partial \rho_{jk}} := \begin{cases}
T_{jk} & \text{if } r = j, s = k, \\
-\frac{1}{l_{jj}} l_{jk} \cdot T_{jk} & \text{if } r = j, s = k,  \\
\sum\limits_{l=0}^{s-1} \frac{\partial L_{rs}}{\partial \rho_{rl}} \cdot \frac{\partial L_{rl}}{\partial \rho_{jk}} & \text{if } r > j, \\
0 & \text{otherwise},
\end{cases}
\end{align}

where $T_{jk}$ arises from the transformation applied in constructing row $j$, and the third case expresses recursive propagation through later rows.
Due to the complexity of implementing this gradient, I used a relatively computationally cheap finite differences approximation of these gradients.
\begin{lstlisting}
#' R Code
#' @param m focal element of vec(Phi_cor)
#' @param Phi_cor_vec vector of partial correlations
#' @param yi observed response vector for person n
#' @param Mu model implied means of observed variables for person n 
#' @param Sigma model implied observed variable covariance matrix
#' @param Lambda model implied factor loading matrix
#' @param Theta model implied residual covariance matrix
#' @param D diagonol matrix of implied factor variances
#' @param eps (default: 1e-6) epsilon value used in finite difference approximation 
finite_difference_partial_corr <- function(m, Phi_cor_vec, yi, Mu, Sigma, Lambda, Theta, D, eps=1e-6){
  M = ncol(Lambda)
  Phi_cor_vec_eps = Phi_cor_vec
  Phi_cor_vec_eps[m] = Phi_cor_vec_eps[m] + eps
  Leps = cholesky_from_partial_corrs(Phi_cor_vec_eps, M)
  Sigma_eps = Lambda %*% Leps %*% D %*% t(Leps) %*% t(Lambda) + Theta
  dgx = (as.numeric(dmvnrm(yi,  Mu, Sigma_eps, logd = TRUE)) - as.numeric(dmvnrm(yi,  Mu, Sigma, logd = TRUE)))/eps
  dgx
}
\end{lstlisting}

Now for the gradients with respect to the moderation effects, $\bm\Delta$.
The person-wise contribution to the vector of gradients with respect to the moderation effects ($\mathbf{x}^t$) are quite straightforward once the above mentioned gradient contributions for each baseline parameter are obtained.
The form of the gradient w.r.t. for a moderating factor $\delta_j$ on any parameter $\theta$ (not to be confused with the item residual variance because here, $\theta$ is a placeholder for any parameter) is thus
\begin{equation}
\frac{\partial L}{\partial \bm\delta_{\theta}} = x_j \cdot \frac{\partial L}{\partial \theta}.
\end{equation}
So the gradient with respect to design variable $x_j$ on item intercept $i$ is therefore
\begin{equation}
\frac{\partial L}{\partial \delta_{\nu_i j}} = x_j\cdot \frac{\partial L}{\partial \nu_i}
\end{equation}
The other gradient contributions are found similarly.
Another important consideration is the approximation of the hessian for approximation of standard errors.

\subsubsection*{Standard errors and the Hessian approximation}

I recommend using a Huber-style sandwich-like estimator for the standard errors using the inverse of the observed information matrix (Hessian) and the outer-product of the gradient at the maximum of the penalized likelihood \citep{Freedman2006-zt}.
Obtaining an analytical definition for the Hessian is not straightforward, while approximating the Hessian has many existing (but slow) implementations. 
The Hessian, or observed information matrix, of the parameter vector, $\bm\theta$, is a necessary contribution to approximate the uncertainty in the estimated parameters.
However, approximating this matrix is not straightforward, and though computationally intensive, the most straightforward approach is to use build in tools to approximate the hessian using numerical methods building on the analytical gradients described above. 

\subsubsection*{Why I abandoned PML for MNLFA}

The MNLFA model was incredibly slow to estimate still for larger sample sizes (N > 10,000).
I kept fiddling with the settings and methods, and I could not find a straightforward way to get over this hurdle. 
This problem only became exponentially worse when I tried to implement a version for item factor analysis.
Especially when it comes to estimating the information function to obtain sound inference and identification of DIF/impact. 

Though, I think the exercise of developing the PML out as I have may be more fruitful if the computational limitations I've mentioned can be overcome with analytic calculation of the information matrix.
The code I wrote for this portion of the project is available on GitHub: \url{https://github.com/noah-padgett/MNLFAcpp}.

\textit{In the next version of this white-paper, I aim to include a novel estimation method I have been working on for these types of models.}


\section*{Discussion}

The penalized maximum likelihood framework for multidimensional moderated nonlinear factor analysis (MNLFA) developed in this paper represents an important step toward addressing the long-standing limitations of existing methods for studying measurement invariance. Traditional applications of MNLFA have been restricted to single- or two-factor models, primarily because of the mathematical and computational difficulty of ensuring valid correlation structures when moderation is introduced at the factor covariance level. By leveraging penalization strategies alongside analytic gradients of the likelihood, our framework allows estimation of models with three or more latent factors in a way that balances flexibility with computational feasibility.

This work builds directly on the original MNLFA framework introduced by Curran and colleagues, who developed a moderated nonlinear factor model for integrative data analysis (IDA) in pooled longitudinal and developmental datasets \citep{Curran2014-jm}. That seminal work highlighted the potential of moderated factor models to accommodate construct non-invariance across heterogeneous subpopulations, but also revealed the challenges of extending beyond two factors. Subsequent research has expanded the methodological foundation of MNLFA, particularly through the adoption of Bayesian approaches. For instance, \citet{Brandt2023-zx, Brandt2025-aa} proposed Bayesian penalty methods using shrinkage priors (e.g., spike-and-slab, lasso-type, and horseshoe priors) to regularize moderation effects and detect differential item functioning (DIF) across multiple moderators. These Bayesian approaches are powerful, especially for simultaneously handling many potential sources of non-invariance, but they are often computationally demanding and require expertise in specifying priors and interpreting posterior distributions.

Our contribution complements this Bayesian line of work by providing a likelihood-based analogue that is both familiar to applied researchers and computationally efficient. In contrast to simulation-based Bayesian estimation, the penalized maximum likelihood framework exploits closed-form gradients of the multivariate normal likelihood with respect to all model parameters, including factor loadings, intercepts, residual variances, factor means, and moderated correlations. This analytic treatment avoids the computational bottleneck of gradient approximation that plagues Bayesian software such as JAGS or Stan. As a result, penalized maximum likelihood estimation is considerably faster, making multidimensional MNLFA accessible for large-scale applications where Bayesian estimation might be prohibitively slow.

The idea of incorporating penalties into factor analytic frameworks is not new, and our work is situated within a growing literature on penalized likelihood methods in psychometrics and structural equation modeling. Finch (2018), for example, demonstrated via Monte Carlo simulations that penalized likelihood estimation improves the detection of measurement non-invariance by controlling Type I error while maintaining or improving power relative to traditional multi-group CFA tests. \citet{geminiani2021penalized} extended penalized factor analysis to both single- and multi-group settings, introducing a general framework for sparsity-inducing penalties in confirmatory models. \citet{hirose2012nonconcave} and \citet{choi2010penalized} developed related non-concave penalized likelihood methods for inducing sparsity in factor loadings, showing how penalization can directly enhance interpretability. Together, this body of work demonstrates that penalties are not only computationally stabilizing but also substantively meaningful, enabling more parsimonious and interpretable representations of latent constructs. Our alignment, ridge, and lasso penalties extend these ideas to the moderation context, offering applied researchers flexible options for calibrating the degree of invariance across model parameters.

Beyond theoretical contribution, penalized MNLFA has significant practical implications. Many substantive domains—including psychopathology, education, and public health—routinely study constructs that are multidimensional, often spanning three to five latent factors. Applied researchers are rarely interested in strict invariance across all measurement parameters, since such invariance is both unrealistic and unnecessary. Instead, they seek to identify where parameters differ across groups or person-level moderators, and to interpret the substantive meaning of such differences. Penalization provides a principled way to detect partial invariance without overfitting or producing unstable estimates. The alignment penalty, for example, encourages approximate invariance across parameters while still allowing localized deviations, echoing the goals of alignment optimization methods in multiple-group CFA \citep{Asparouhov2024-ia}.

Despite these advantages, challenges remain. A primary computational bottleneck arises from the need to construct person-specific implied means and covariance matrices in the likelihood evaluation. Because moderation enters directly into item and factor parameters, sufficient statistics cannot be used to simplify the likelihood, as is possible in conventional CFA software such as lavaan. Instead, the density function must be evaluated separately for each individual, limiting scalability. While our derivations of analytic gradients alleviate much of this burden, further efficiency gains may come from algorithmic innovations such as caching strategies, low-rank approximations, or GPU acceleration. This echoes concerns raised by \citet{ludtke2021penalized} , who found that small-sample confirmatory factor analysis often struggles with convergence under MCMC, but can benefit from penalized ML approaches due to their relative stability.

Another area for development is the design of penalty functions themselves. While I focus here on ridge, lasso, and alignment penalties, other formulations may offer distinct advantages. Elastic net penalties, for example, combine the strengths of ridge and lasso by balancing smooth shrinkage with sparsity induction. Adaptive penalties can weight moderation effects differently, improving bias-variance tradeoffs. Group penalties could be used to enforce structure across related sets of moderation effects (e.g., all loadings associated with a particular factor). Systematic simulation studies will be essential to compare the performance of these alternatives under varying conditions of dimensionality, sample size, and degree of true non-invariance.

Extending penalized MNLFA beyond continuous indicators also represents a promising direction. Many applied settings involve binary or ordinal indicators, longitudinal repeated measures, or mixtures of latent subgroups. While the general penalized likelihood framework extends naturally to such cases, additional derivations are required for closed-form gradients in non-Gaussian models. Software implementations are another key frontier. Current packages such as aMNLFA \citep{Gottfredson-2019} and mnlfa \citep{robitsch2023mnlfa} have increased accessibility of MNLFA by automating score generation and incorporating penalization for limited-factor models. Developing similar tools for multidimensional penalized MNLFA will be critical for broader adoption. Integration into widely used SEM platforms or stand-alone R packages could help disseminate these methods to applied researchers who might otherwise be deterred by computational or coding complexity.

In conclusion, the penalized maximum likelihood framework presented here advances the methodological toolkit for evaluating measurement invariance in multidimensional constructs. It extends the seminal contributions of \citet{Curran2014-jm} and complements recent Bayesian penalty approaches \citep{Brandt2025-aa} by offering a frequentist alternative that is both interpretable and computationally efficient. By combining analytic gradients with penalization, our approach makes multidimensional MNLFA practically estimable and theoretically robust. Challenges remain in scalability, penalty design, and software implementation, but the path forward is clear: continued development, comparative validation, and applied demonstration will help realize the full potential of penalized MNLFA for promoting fairness, validity, and precision in measurement across diverse populations.




\bibliographystyle{apalike}
\bibliography{multidimensional_mnlfa}

\end{document}